\documentclass[useAMS,usenatbib]{mn2e}

\usepackage{graphicx}

\newcommand{\swift}{{\it Swift}}
\newcommand{\xmm}{{\it XMM-Newton}}

\newcommand{\kepler}{{\it Kepler}}

\title[PDS of \kepler\ observations of four AGNs]{New structures of Power Density Spectra for four \kepler\ AGNs}
\author[A. Dobrotka et al.]{
A. Dobrotka$^1$\thanks{E-mail: andrej.dobrotka@stuba.sk}, V. Antonuccio-Delogu$^{2}$ and I. Baj\v{c}i\v{c}\'akov\'a$^{1}$\\
$^1$Advanced Technologies Research Institute, Faculty of Materials Science and Technology in Trnava, Slovak University of Technology\\
in Bratislava, Bottova 25, 917 24 Trnava, Slovakia\\
$^{2}$INAF-Catania Astrophysical Observatory, Via S. Sofia 78, I-95123 Catania, ITALY\\
}

\begin{document}

\date{Accepted ???. Received ???; in original form \today}

\pagerange{\pageref{firstpage}--\pageref{lastpage}} \pubyear{2017}

\maketitle

\label{firstpage}

\begin{abstract}
Many nearby AGNs display a significant short-term variability. In this work we re-analyze photometric data of four active galactic nuclei observed by \kepler\ in order to study the flickering activity, having as main goal that of searching  for multiple components in the power density spectra. We find that all four objects have similar characteristics, with two break frequencies at approximately log($f$/Hz)=-5.2 and -4.7. We consider some physical phenomena whose characteristic time-scales are consistent with those observed, in particular mass accretion fluctuations in the inner geometrically thick disc (hot X-ray corona) and unstable relativistic Rayleigh-Taylor modes. The former is supported by detection of the same break frequencies in the \swift\ X-ray data of ZW229-15. We also discuss rms-flux relations, and we detect a possible typical linear trend at lower flux levels. Our findings support the hypothesis of a multiplicative character of  variability, in agreement with the propagating accretion fluctuation model.
\end{abstract}


\section{Introduction}
\label{introduction}

Active galactic nuclei (AGNs) are distant galaxies hosting supermassive black holes (BHs) as the main engine driving the accretion process at their centers (see e.g. \citealt{peterson1997} for a review). Similar to the case of interacting binaries, the accretion process generates very typical radiative finger-prints in the form of "fast" (i.e. few days or less) stochastic variability, also called flickering (see e.g. \citealt{warner1995}, \citealt{lewin2010} for a review). In general these objects are assumed to possess an accretion disc (except for binaries with strong magnetic fields) which drives the matter from the periphery to the central compact object.

Different disc structures, interactions and instabilities are active during the mass flow, which makes the study of the accretion process very complex. Each structure or physical process is characterized by its own time-scale. The very basic and promising explanation of flickering in interacting binaries is the propagation of accretion fluctuations in an unstable accretion flow (\citealt{lyubarskii1997}, \citealt{king2004}, \citealt{arevalo2006}). For these viscous process a general scaling rule is that characteristic time-scale scales with BH mass. Therefore, the study of flickering is highly demanding both in terms of temporal extent and observational quality. The duration of an observation limits the largest detected time-scale, while time sampling determines the shortest time-scale. Both these requirements have been  to an unprecedented quality by the \kepler\ mission (\citealt{borucki2010}). This mission, whose primary goal is the search for extra-solar planets, has also observed different objects in the field, like AGNs. The typical light curve cadence is of $\sim 30$\,min and the total duration of the observation reaches few hundreds of days. These features allow one to study the fast variability of objects like AGNs with details not attainable from the ground.

The statistical characteristics of flickering carries many physical information about the source. One of the fundamental finger-prints of the underlying accretion process is the linear relation between the variability amplitude and the mean flux (\citealt{uttley2001}, \citealt{uttley2005}). The variability amplitude (usually called rms) is defined as the square root of the variance and the typical linear relation suggests that the variability is coupled multiplicatively and can be explained by variations in the accretion rate that are produced at different disc radii and propagate inward (\citealt{lyubarskii1997}, \citealt{kotov2001}, \citealt{arevalo2006}). The relation is observed in a variety of accreting systems as X-ray binaries, active galactic nuclei (\citealt{uttley2005}), cataclysmic variables (CVs) stellar systems (\citealt{scaringi2012a}, \citealt{vandesande2015}) and symbiotic systems (\citealt{zamanov2015}). There exist however also systems deviating from this linear relationship for which we have evidence of flickering caused by fluctuations of the accretion rate. \citet{dobrotka2016a} showed that if a variability with linear rms-flux trend is superposed on independent variability with longer time-scale, the rms-flux trend deviates from linearity.

Another frequently observed feature of flickering evidenced in a log-log periodogram (the so called power density spectrum, PDS) is a red noise, which shows itself as a linear relationship between power density and frequency. Most often, red noise is detected in some frequency interval, and its value changes near a specific characteristic break frequency. The latter is a typical finger-print of the underlying physical processes.

The PDSs of interacting binaries show multiple components (see e.g. \citealt{scaringi2012b}, \citealt{dobrotka2016a}) generated by different components of the accretion flow as accretion disc, central evaporated hot corona etc. It is suggestive to expect the same also in the case of AGNs. The first study of AGNs observed by \kepler\ was performed by \citet{mushotzky2011}. These authors did not detect any break frequency in the PDSs and concluded that the slopes are very steep, which is typical for the longer time-scales compared to the interacting binaries usually showing much shallower red noises. Subsequent re-analysis of ZW229-15 \kepler\ data by \citet{edelson2014} revealed a characteristic break frequency suggesting the existence of time-scale of $\sim 5$\,day in the PDS. The latter authors concluded that the viscous time-scale is not consistent with these findings. Dynamical or thermal processes or time crossing phenomena are more probable physical processes behind them.

The discovery of multi-component PDSs in interacting binaries has been gradual. Ground based observations first showed a simple broken power law (see e.g. \citealt{kato2002}, \citealt{baptista2008}), i.e. red noise with a single break frequency. After the appearance of instruments like \kepler, more components were discovered in the PDSs. But re-analysis of old data using different methods resulted into the discovery of more structured PDSs (see e.g. \citealt{dobrotka2014}). This PDSs shape upgrade in interacting binaries together with the re-analysis of the ZW229-15 data by \citet{edelson2014} and the discovery of a more complex PDS structure motivated us to revisit all four AGNs already studied by \citet{mushotzky2011}, searching for multiple-components in the PDS. We use the same method we used for CVs studies (see e.g. \citealt{dobrotka2014}, \citealt{dobrotka2016a}) and we shortly present the method in Section~\ref{method}. In the subsequent Section~\ref{fits} we present the identification of PDS components, and in Section~\ref{simulations} we present some additional simulations motivated by the shape of the light curve assembly. In Section~\ref{rms-flux} we present an analysis of the rms-flux data, and finally we discuss our finding in Section~\ref{discussion}.

\section{Data}

\subsection{Kepler}

The NASA \kepler\ mission (\citealt{borucki2010}) offers a unique opportunity to study PDSs of a variety of accreting objects over a wide frequency range, thanks to the high quality of the data: low noise, without any perturbation from the atmosphere, and almost continuous coverage during long time with a cadence of 30\,min. This continuity is only perturbed by periodical (every third month) $90\deg$ rolls needed to keep the solar panels oriented towards the Sun. Otherwise, the light curves can have a duration up to 1400\,days (\citealt{dobrotka2016a}). Further monitoring of the objects reaching even longer light curve has been disabled due to a breakdown of a second reaction wheel in 2013.

We downloaded standard reduced light curves of four AGNs observed by \kepler, i.e. ZW229-15 (\kepler\ ID: 6932990), KA1925+50 (\kepler\ ID: 12158940), KA1904+37 (\kepler\ ID: 2694186) and KA1858+48 (\kepler\ ID: 11178007), all already studied by \citet{mushotzky2011}. The light curves are partitioned into several subsamples which we compose together to get a single light curve. Since the subsamples have different flux levels\footnote{A simple subsample addition does not result in a continuous light curve, and the subsamples are scattered vertically.}, we match the last point of one curve with the first point of the subsequent one. This procedure is very simple, but in Section~\ref{simulations} we show that this rough flux correction does not influence the results, thus more sophisticated composition methods are not necessary. In some cases the observational gap includes also the subsample division, then keeping the same flux level is not realistic. To understand the impact of these light curves artifacts on our analysis we performed some test simulations (see Section~\ref{simulations}).

\begin{figure}
\resizebox{\hsize}{!}{\includegraphics[angle=-90]{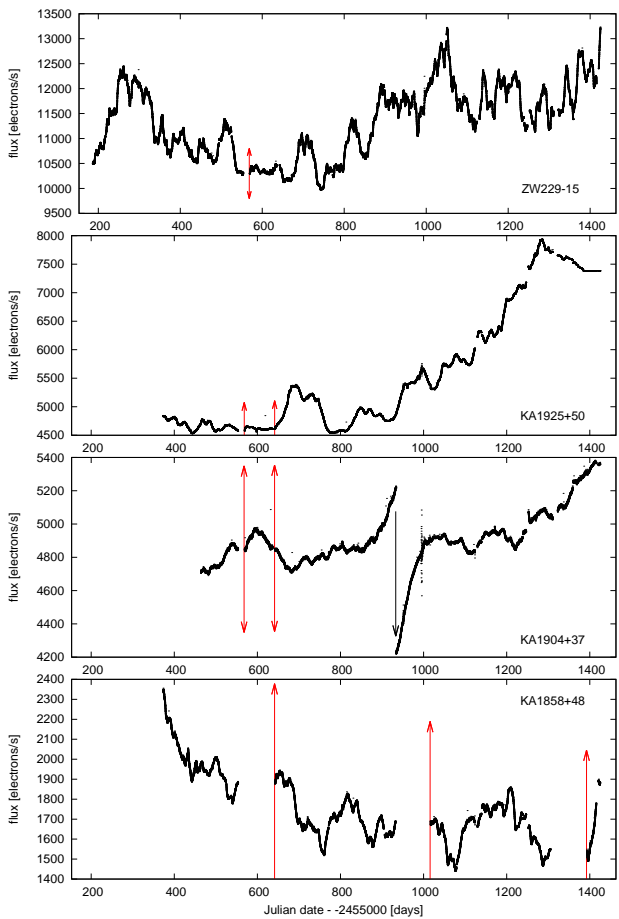}}
\caption{Individual light curves of the analysed AGNs. The curve for KA1904+37 is divided into two parts, with vertical offset by -1000 since day 1100 (marked by black arrow) for visualization purposes. The red arrows show the simulated flux scatter in controversial observational gaps (see Section~\ref{simulations} for details).}
\label{lc_all}
\end{figure}

The light curves are presented in Fig.~\ref{lc_all}. The object KA1904+37 experienced a strong brightness increase during the observation, therefore we divided the light curve in two parts. We have offset the second half of the light curve by -1000 (starting from day 1100) for better visualization of the fast variability.

\subsection{SWIFT}

\begin{figure}
\resizebox{\hsize}{!}{\includegraphics[angle=-90]{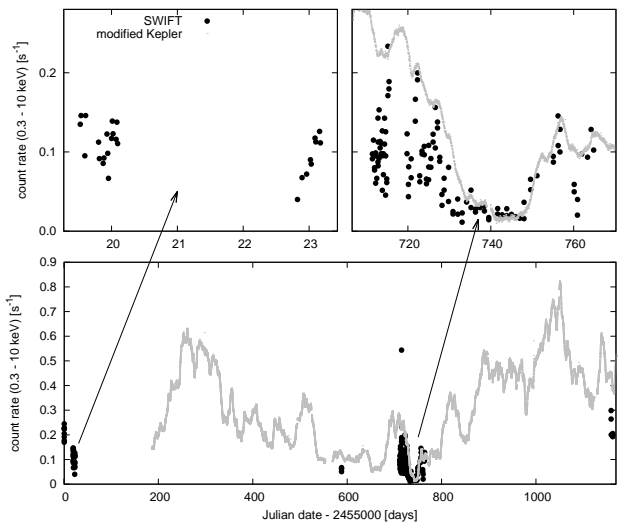}}
\caption{\swift\ light curve of ZW229-15 (black thick points). Two zoomed regions are shown in the upper panels while the whole data set is in the lower panel. The arrows localize the zoomed light curve subsamples in the whole data set. Vertically modified Kepler data (gray dots) are shown for comparison.}
\label{lc_swift}
\end{figure}

For one of our AGN (ZW229-15) public data from \swift\ satellite are publicly available. The X-ray light curve\footnote{Taken from http://www.swift.ac.uk/user\_objects/} in a band 0.3 - 10\,keV from the XRT instrument with 1000\,s binning is presented in Fig.~\ref{lc_swift}. \kepler\ data are added for comparison, with the flux modified as flux/4000+0.14 to make the data consistent and directly comparable.

\section{PDS analysis}

\subsection{Method}
\label{method}

To compute the PDS we used the Lomb-Scargle algorithm (\citealt{scargle1982}), which can handle gaps in the light curves and/or non equidistant data. We first divide the light curves into 5 subsegments\footnote{Our choice of the number of subsamples is empirically based on visual inspection of the PDS, aiming at choosing an adequate model for PDS fitting.} of equal length. Then we calculate a log-log periodogram for each subsample. All five periodograms are then combined to get a mean periodogram. We perform an additional binning with constant width of 0.05 to finally obtain a Power Density Spectrum estimate\footnote{What we describe here is rather an averaged and binned periodogram. The real power density spectrum is without the inherently present scatter. However, the averaged and binned case produces a result very close to it.} (PDS). As an indicator of the intrinsic scatter within each averaged frequency bin we use the standard error of the mean from all points. The choices of the frequency interval and step depend on the characteristics of the subsample. The low frequency end and the frequency step of the PDSs usually are chosen according to the temporal extent of the  light curve . The high frequency end is limited by white noise, and we empirically defined this limit to be: log($f/{\rm Hz}) = -4.0$. As y-axis units we use frequency multiplied by power ($f \times p$).

\begin{figure}
\resizebox{\hsize}{!}{\includegraphics[angle=-90]{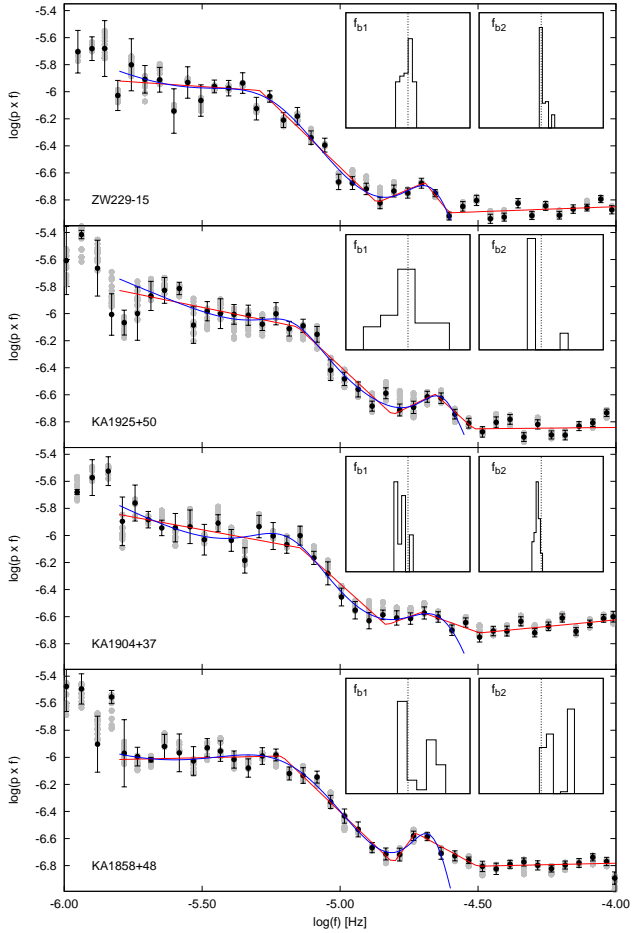}}
\caption{Individual observed PDSs (black points with error bars) with multi-component broken power (red line) law and Lorentzian (blue line) fits. The inset panels show the histograms of fitted frequencies $f_{\rm b1}$ and $f_{\rm b2}$ after 100 simulations (corresponding PDS points depicted as gray points) with different light curve configurations (see Section~\ref{simulations} for details). The vertical dotted line is the mean value from Table~\ref{pds_param} and the x-axis interval is the corresponding error interval.}
\label{pds_all}
\end{figure}

We present the resulting PDSs from log($f/{\rm Hz}) = -6.0$ in Fig.~\ref{pds_all}. We omitted lower frequencies because the scatter is too large and hinders a consistent analysis. The power is rising toward lower frequencies. The multi-component shape is visible, with a lower break frequency ($f_{\rm b1}$) around ${\rm log}(f/{\rm Hz}) = -5.2$ and higher frequency component ($f_{\rm b2}$) around ${\rm log}(f/{\rm Hz}) = -4.7$ in all four cases.

\subsection{Fitted models}
\label{fits}

For the PDS fitting we used two different models using {\small GNUPLOT}\footnote{http://www.gnuplot.info/} software. The first model is a broken power law model consisting of linear functions, where individual characteristic break frequencies are the joining points between two consecutive linear functions. The second model is a combination of a Lorentzian profile and of an additional function representing the underlying rising PDS trend toward lower frequencies (see e.g. \citealt{scaringi2014}). The Lorentzian has a form of
\begin{equation}
p(f) = \frac{a \Delta}{\pi}\frac{1}{\Delta^2 + (f - f_1)^2},
\end{equation}
where $p$ is the power in units of $f \times p$, $a$ is the normalization free parameter, $f$ is the frequency, $f_1$ is the searched break frequency and $\Delta$ is the half-width at half-maximum. In order to fit the rising trend towards lower frequencies we choose a function of the form:
\begin{equation}
p(f) = \frac{A f^{-1}}{1 + f/f_2},
\end{equation}
where $A$ is a normalization and $f_2$ is the break frequency. With this definition the PDS shape has a slope $-1$ below $f_2$ and $-2$ above $f_2$. For the Lorentzian model we chose different higher frequency PDS limits, because the model does not include the white noise. This end is defined visually by the higher PDS component $f_{\rm b2}$. 

\begin{figure}
\resizebox{\hsize}{!}{\includegraphics[angle=-90]{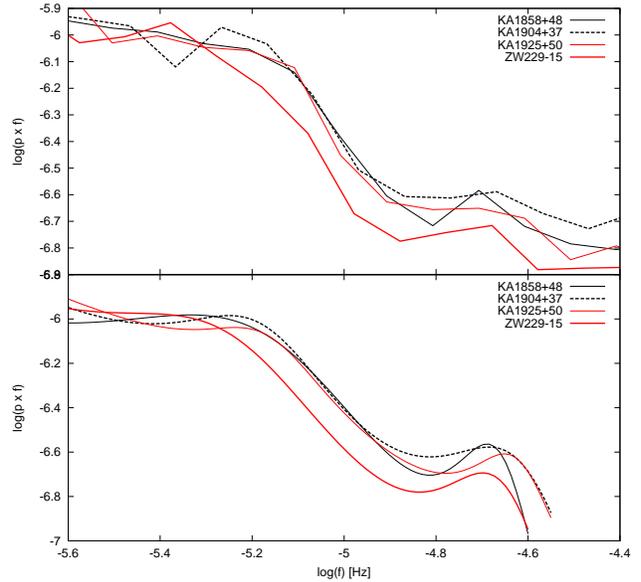}}
\caption{Direct comparison of PDSs (upper panel) and multi-component Lorentzian fits (lower panel) from Fig.~\ref{pds_all}. The PDSs in the upper panel are depicted with double binning (bin interval is 0.1) than in Fig.~\ref{pds_all} to reduce the scatter.}
\label{pds_comp}
\end{figure}

We performed the fits over a selected frequency interval within which the inherent scatter is sufficiently small: this results in a value of ${\rm log}(f/{\rm Hz}) = -5.8$ as a lower PDS frequency end. Individual fits are presented in Fig.~\ref{pds_all} and fitted parameters listed in Table~\ref{pds_param}. Fig.~\ref{pds_comp} shows direct comparison of the PDSs and the fitted Lorentzian models.
\begin{table}
\caption{PDS parameters as lower ($f_{\rm b1}$) and higher ($f_{\rm b2}$) break frequency with standard errors (calculated from the covariance matrix by the {\small GNUPLOT} software) from different models. Reduced $\chi^2$ and degree of freedom (dof) are listed too. The upper and lower values for every object are from multi-component broken power law and multi-component Lorentzian models, respectively. Parameters obtained from \swift\ data using broken power law fit are in parenthesis.}
\begin{center}
\begin{tabular}{lcccc}
\hline
\hline
object & log($f_{\rm b1}/{\rm Hz}$) & log($f_{\rm b2}/{\rm Hz}$) & $\chi^2_{\rm red}$ & dof\\
\hline
ZW229-15 & $-5.29 \pm 0.04$ & $-4.69 \pm 0.49$ & 1.64 & 26\\
 & $-5.24 \pm 0.08$ & $-4.65 \pm 0.11$ & 0.99 & 15\\
 & ($-5.48 \pm 0.20$) & ($-4.65 \pm 0.15$) & (3.79) & (20)\\
KA1925+50 & $-5.15 \pm 0.07$ & $-4.66 \pm 0.05$ & 3.90 & 27\\
 & $-5.17 \pm 0.04$ & $-4.62 \pm 0.05$ & 2.02 & 16\\
KA1904+37 & $-5.14 \pm 0.05$ & $-4.70 \pm 0.07$ & 1.29 & 27\\
 & $-5.18 \pm 0.05$ & $-4.61 \pm 0.16$ & 0.79 & 16\\
KA1858+48 & $-5.21 \pm 0.02$ & $-4.73 \pm 0.01$ & 0.97 & 27\\
 & $-5.15 \pm 0.14$ & $-4.66 \pm 0.05$ & 0.86 & 15\\
\hline
\end{tabular}
\end{center}
\label{pds_param}
\end{table}

\begin{figure}
\resizebox{\hsize}{!}{\includegraphics[angle=-90]{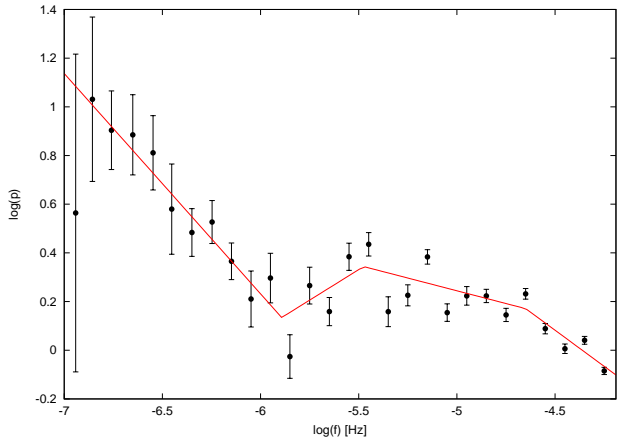}}
\caption{PDS of ZW229-15 calculated from X-ray \swift\ data with multi-component broken-power law fit.}
\label{pds_swift}
\end{figure}

PDS calculated from X-ray \swift\ data is shown in Fig.~\ref{pds_swift}. Due to low quality and quantity of data we have not further decomposed the light curve into subsamples.  We compute a single periodogram and we apply a binning to estimate the PDS. For visualization purposes we use different y-axis units than for the \kepler\ data (Lomb-Scarge power) due to fact that we find a very shallow PDS. Only a broken power law fit yields acceptable result. In the case of multi-component Lorentzian model, only one component converged to a value with the studied frequency interval, i.e. ${\rm log}(f/{\rm Hz}) = -5.17 \pm 0.91$. Resulting break frequencies from multi-component broken-power law best-fitting are summarized in Table~\ref{pds_param}.

\subsection{Tests of light curves reliability}
\label{simulations}

When constructing the light curves, we found that some observational gaps in the temporal sequences almost coincided with the subsample boundaries. In some particularly controversial cases (marked as red arrows in Fig.~\ref{lc_all}), where the gap is too wide, it is impossible to adjust the flux in order to get a continuous light curve. As we wrote before, in these cases an ad-hoc solution where the final and initial points of two adjacent subsamples are forced to coincide was adopted to construct the light curve. In order to test the influence of this particular choice on our results, we performed 100 simulations were we uniformly scattered the flux offset from -500 to 500 (the length of the red arrows in Fig.~\ref{lc_all} corresponds to this scatter on the y-axis). Apparently, the scatter is too large and even overestimated in KA1904+37 and KA1858+48, while in the two other cases it represents the observed flux variability.

We show the resulting PDSs as gray points in Fig.~\ref{pds_all}. The statistics of the fitted PDS parameters $f_{\rm b1}$ and $f_{\rm b2}$ are shown in the inset panels of Fig.~\ref{pds_all}. The fitted break frequencies in simulations all lie within the errors of the original values. Moreover, the lower the frequency, the higher the scatter of the simulated values (gray points). Therefore, the flux uncertainty mainly affects the lower frequency part of the PDS, as intuitively expected. Thus our interval of interest is not affected considerably and the fitted results are robust against the light curve construction procedure within the simulated flux scatter. Larger scatter values are unlikely.

\section{Rms-flux analysis}
\label{rms-flux}

\begin{figure}
\resizebox{\hsize}{!}{\includegraphics[angle=-90]{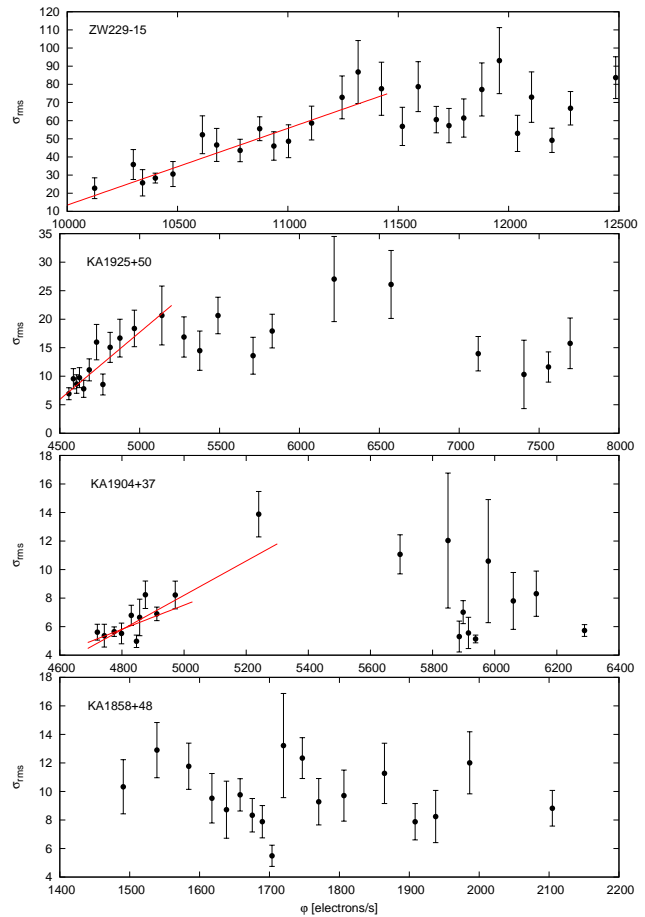}}
\caption{Low resolution rms-flux data from \kepler\ light curves with linear fits shown as red solid lines. Error bars are standard errors of the mean within each bin. In the case of KA1904+37 two different fits performed over two different flux intervals are shown.}
\label{rms-flux_large}
\end{figure}

The absolute rms amplitude of variability is defined as square-root of the variance, i.e.
\begin{equation}
\sigma_{\rm rms} = \sqrt{\frac{1}{N - 1} \sum^N_{i = 1} (\psi_i - \overline{\psi})^2},
\end{equation}
where $N$ is the number of observed fluxes $\psi_i$ and $\overline{\psi}$ is their mean value. For $\sigma_{\rm rms}$ and $\overline{\psi}$ calculation we used 200\footnote{The value $N=200$ was chosen on empirical grounds: smaller values yield just more scattered rms-flux data. Similar results can be achieved by first binning the original light curve and computing the rms-flux using smaller number of points.} consecutive points and we binned the data ($\sigma_{\rm rms}$,$\overline{\psi}$) in groups of 10. We present the resulting rms-flux data in Fig.~\ref{rms-flux_large}. All objects except KA1858+48 show possible linear ascending trends for the lowest fluxes. The most pronounced linearity is in ZW229-15. In all cases an initial ascending trends is followed by a rms plateau or a slight depression at higher values of the flux. The linear fits are depicted as red solid lines.

\begin{figure}
\resizebox{\hsize}{!}{\includegraphics[angle=-90]{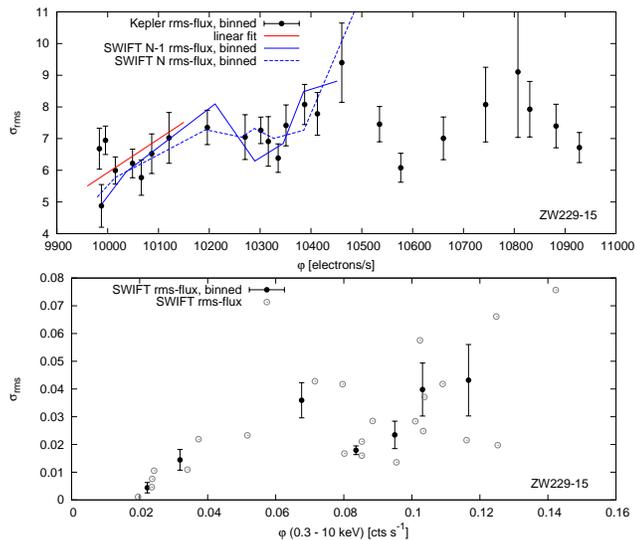}}
\caption{Same as Fig.~\ref{rms-flux_large} but with higher flux resolution (10 \kepler\ and 5 \swift\ light curve points are used for $\sigma_{\rm rms}$ and $\overline{\psi}$ calculation) and for restricted time interval of \swift\ data from upper right panel of Fig.~\ref{lc_swift}. The \swift\ rms-flux data in the lower panel were calculated with the anomalous very high count rate point (close to day 720 in the lower panel of Fig.~\ref{lc_swift}) excluded. The binned \swift\ data (two versions using all N rms-flux points and N-1 points) as blue lines are added into upper panel and horizontally/vertically modified for direct comparison.}
\label{rms-flux_short}
\end{figure}

In Fig.~\ref{rms-flux_short}, we show the rms-flux data calculated from the \swift\ and \kepler\ data on the same time interval (from upper right panel of Fig.~\ref{lc_swift}). In this case only ten \kepler\ light curve points were used for $\sigma_{\rm rms}$ and $\overline{\psi}$ calculation, and points were binned with each bin containing ten data points. For \swift\ data we used only 5 points for $\sigma_{\rm rms}$ and $\overline{\psi}$ calculation and every 3 ($\sigma_{\rm rms}$,$\overline{\psi}$) points were binned because of the very low number of \swift\ data\footnote{Note that we used two data sets. One with all data, and second without the one anomalous point (visible in lower panel of Fig.~\ref{lc_swift} close to day 720) excluded.}. We also show in the upper panel the linear fit as a red solid line and the \swift\ rms-flux binned data as blue lines, for a direct comparison. The latter has been adapted horizontally and vertically to find a match. Apparently both data sets show an initial rise followed by a plateau or by a slight depression, with a subsequent ascension. A comparison between the \kepler\ and \swift\ figures hints at a possible correlation between optical and X-ray variability.

\section{Discussion}
\label{discussion}

In this paper we have re-analyzed optical light curves of four AGNs observed by \kepler\ and we have discovered multiple component in their PDSs, with evidence for two break frequencies centered around ${\rm log}(f_{\rm b1}/{\rm Hz}) = -5.2$ and ${\rm log}(f_{\rm b2}/{\rm Hz}) = -4.7$. A value close to the former had already been detected by \citet{edelson2014} for ZW229-15. The authors derived a value of $0.18 \pm 0.03$\,day$^{-1}$ which transforms to ${\rm log}(f/{\rm Hz}) = -5.68^{+0.07}_{-0.08}$.

\subsection{Break frequencies in the ZW229-15 \kepler\ PDS}

\begin{figure}
\resizebox{\hsize}{!}{\includegraphics[angle=-90]{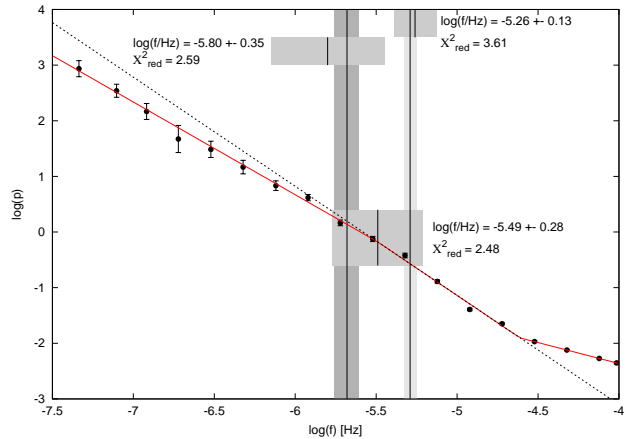}}
\caption{Low resolution PDS (different binning than in Fig.~\ref{pds_all}) version of ZW229-15. The red line shows the best broken power law fit with two break frequencies. The black dotted line is the extrapolated middle power law for comparison. Vertical lines with the shaded areas are the frequency values with uncertainty; darkest is the value derived by \citet{edelson2014} and the brightest is the value derived in this paper. The three boxes show three fitted break frequencies with uncertainties and marked $\chi^2_{\rm red}$ of the corresponding fit using different initial parameter estimates (see text for details).}
\label{pds_low_res}
\end{figure}

The two break frequencies we have found do not agree within the errors, suggesting that we found PDS components different from those found by \citet{edelson2014}. This statement however must be proved and to that purpose we performed a different analysis, similar to that of \citet{edelson2014}. First, we detrended the light curve with a linear function to have the average flux from the first 20 points of each light curve equal to the average of the last 20 points. Second, we did not divide the light curve of ZW229-15 into several subsamples, i.e. we used all the data points as a whole. Third, as dependent variable we just adopted power. Finally, we used a larger frequency bin to construct the PDS (interval of 0.2). Fig.~\ref{pds_low_res} shows the PDS together with the best-fitting using multi-component power law fit (3 components, red curve). The value of the break frequency converged to three values, slightly depending on the starting initial parameter, i.e. $-5.49 \pm 0.28$\,Hz ($\chi^2_{\rm red} = 2.48$), $-5.80 \pm 0.35$\,Hz ($\chi^2_{\rm red} = 2.59$), $-5.26 \pm 0.13$\,Hz ($\chi^2_{\rm red} = 3.61$). The worst fit agrees with the frequency $f_{\rm b1}$ derived in this paper, while the best value agrees with the one derived by \citet{edelson2014}. The frequency values with error intervals are shown as shaded areas in Fig.~\ref{pds_low_res}. While a smooth bent model as used by \citet{edelson2014} would be more adequate, we kept the test to be equivalent to the analysis in Section~\ref{fits}. Reducing the binning interval to 0.1 yields similar results, but with a worse $\chi^2_{\rm red}$ due to the larger PDS scatter.

We thus conclude that the break frequency $f_{\rm b1}$ detected in this paper is different from that detected by \citet{edelson2014}. The latter changes the overall trend of the PDS, while the former is a local hump/substructure. The reason why Edelson et al. did not find the frequency $f_{\rm b1}$ has to be searched in the different analysis method (division of the light curve into subsamples), but mainly on different binning because a binning not fine enough, can destroy finer PDS details. The y-axis units may play a role also. While both y-axis units are equivalent for a modeling, the $f \times p$ case is better\footnote{This is an empirical statement. However, we notice that when using $f \times p$ for the y-axis the resulting curves are less steep, therefore reducing the y-axis interval. This results in lower vertical expansion of the PDS, ultimately producing a better visualization of details needed for model selection and initial parameter adjustment before fitting procedure.} for a detailed visualization of details (see e.g. \citealt{scaringi2014}).

\subsection{Time-scale estimates}

\subsubsection{"Classical" case}
\label{classical_case}

\citet{edelson2014} made basic time-scale estimates in order to identify the source of the detected break frequency. Their measured value is close to the value $f_{\rm b1}$ detected in this paper, therefore the conclusion of \citet{edelson2014} may still be valid, although a plausible association could be made with perturbations occurring on light-crossing and dynamical ($t_{\rm dyn}$) time-scales, but not with the viscous time-scale ($t_{\rm visc}$). The exception is the thermal time-scales ($t_{\rm th}$) estimate, which is already too long to be compatible with the time-scale associated to $f_{\rm b1}$.

The three time-scales $t_{\rm dyn}, t_{\rm th}$ and $t_{\rm visc}$ are connected as follows:
\begin{equation}
t_{\rm dyn} \sim \alpha t_{\rm th} \sim \alpha (H/R)^2 t_{\rm visc},
\end{equation}
where $H$ and $R$ are the height scale of the disc and distance from the center, respectively (see e.g. \citealt{king2008}). $t_{\rm dyn}$ is estimated from circular orbits:
\begin{equation}
t_{\rm dyn} \sim \left( \frac{R^3}{GM} \right)^{1/2},
\end{equation}
where $G$ and $M$ are the gravitational constant and BH mass, respectively. Finally, a realistic estimate for $t_{\rm visc}$:
\begin{equation}
t_{\rm visc} \sim \frac{t_{\rm dyn}}{\alpha (H/R)^2}.
\label{eq_tvisc}
\end{equation}

\citet{edelson2014} estimated a disc radius of 100 - 1000 Schwarzschild radii (R$_{\rm schw}$) yielding $t_{\rm visc} \simeq 4 - 140\,$ yrs, for a BH mass $M\approx 10^{7} {\rm M}_{\odot}$. This is not consistent with the detected 5\,day time-scale. The authors concluded that this might be regarded as an evidence against the propagating fluctuation model (\citealt{lyubarskii1997}, \citealt{king2004}, \citealt{arevalo2006}), which is however a very basically grounded model of accretion in interacting binaries. However, \citet{edelson2014} suggested that the whole optically thick, geometrically thin disc could be the source of the variability. Notice that they used $\alpha \sim H/R \sim 0.1$. \citet{scaringi2014} analyzed similar multi-component PDS of the CV MV\,Lyr, and modeled the highest break frequency ${\rm log}(f/{\rm Hz}) = -3$ assuming a model based on propagating fluctuations in the accretion flow. The results are not compatible with a geometrically thin disc, but the derived value of $\alpha (H/R)^2 = 0.705^{+0.289}_{-0.182}$ suggests a $H/R$ ratio higher than approximately 0.75. \citet{scaringi2014} concluded that the source of such variability can not be the standard optically thick and geometrically thin disc, but a central optically thin and geometrically thick disc known also as hot evaporated corona. Mass accretion fluctuations are the variability source of X-rays which are reprocessed at optical wavelengths by the underlying geometrically thin disc.

Such corona can arise as a result of inefficient cooling in the innermost regions of the disc (see e.g. \citealt{meyer1994}). A dwarf nova as a subclass of CVs has two phases: an outburst when the accretion disc reaches down to the white dwarf surface, followed by a quiescent one where the central disc is truncated or evaporated and a hot corona is instead present (see e.g. \citealt{warner1995} for a review). \citet{scaringi2014} proposed a solution of a sandwiched model, where the central corona surrounds the central region of the standard accretion disc reaching the white dwarf surface, i.e. a situation which can be typical for a subclass to which MV\,Lyr belongs. This statement was confirmed by \citet{dobrotka2016b} by direct X-ray observations using the \xmm\ satellite.

We analysed long-term data of ZW229-15 from \swift\ satellite and we conclude that despite the low quality and quantity of data, the presence of both break frequencies is possible in the corresponding X-ray PDS. The lower break frequency of ${\rm log}(f_{\rm b1}/{\rm Hz}) = -5.48$ agrees within the errors with the optical value derived from \kepler\ data, while the higher frequency ${\rm log}(f_{\rm b2}/{\rm Hz}) = -4.65$ matches (almost) perfectly. The $\chi^2_{\rm red}$ of 3.79 suggests a very bad fit, but obviously this is due to large scatter of the PDS around the break frequency. Such low quality PDS is not adequate for reliable break frequency measurements, but it supports the X-ray origin of the fluctuations.

\begin{figure}
\resizebox{\hsize}{!}{\includegraphics[angle=-90]{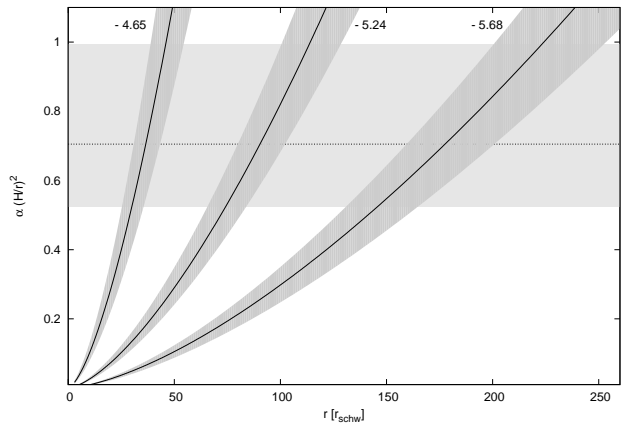}}
\caption{$\alpha (H/R)^2$ values vs. outer radius of a geometrically thick disc assuming viscous processes as the source of variability. Solid lines with the shaded areas are the solutions with uncertainty intervals based on three break frequencies (marked in the top of the figure in log values) derived by \citet{edelson2014} and in this paper for ZW229-15. The horizontal dotted line with the shaded area is the $\alpha (H/R)^2$ value with uncertainty interval derived for similar geometrically thick disc in the CV MV\,Lyr shown for comparison.}
\label{time_scales}
\end{figure}

Therefore, if the source of the variability is the central hot corona, it does not reach the outer radii of the underlying geometrically thin disc and lower values are possible. We took this solution into account and calculated the $\alpha (H/R)^2$ parameter using all three derived break frequencies of ZW229-15. The $\alpha (H/R)^2$ values vs. disc radii $R$ is shown in Fig.~\ref{time_scales}. Apparently a solution is possible for all three break frequencies with the $\alpha (H/R)^2$ values suggesting a geometrically thick disc with outer disc/corona radii up to $\sim 250$\,R$_{\rm schw}$ supposing maximal parameter values $\alpha \sim H/R \sim 1$. If the hot corona in ZW229-15 has similar characteristics as for interacting binaries, or more specifically as that of the CV MV\,Lyr, the outer disc/corona radii is between 26 and 252\,R$_{\rm schw}$ (taking all errors into account). For a geometrically thin disc with $\alpha \sim H/R \sim 0.1$ as used by \citet{edelson2014} the value of $\alpha (H/R)^2$ should be $\sim 0.001$. However, The $H/R$ ration could even be smaller ($\simeq 0.01$) yielding even smaller values of $\alpha (H/R)^2$. Therefore, the geometrically thick disc is the only solution if we suppose a viscous origin of the fluctuations.

An exhaustive analysis aiming at identifying the sources of individual PDS components is beyond the scope of this paper, because it is difficult to make it only using such simplistic time-scale estimates. Direct modeling as performed by \citet{scaringi2014} can however yield significant results. Using the same $t_{\rm visc}$ estimate (equation~\ref{eq_tvisc}) for the CV MV\,Lyr we get a value of 105.5\,s (using outer corona/disc radius, $\alpha (H/R)^2$ parameter and primary mass from \citealt{scaringi2014}), which gives a values of ${\rm log}(f/{\rm Hz}) = -2.02$. The modeled highest break frequency has a value of approximately ${\rm log}(f/{\rm Hz}) = -3$, i.e. an order of magnitude smaller. Although these time-scale estimates are very approximate, we conclude that taking into account the corona as a possible source of the variability, the viscous processes together with the idea of propagating accretion fluctuations represent a viable physical explanation. Besides a modeling, X-ray observations can give a final answer, which would be even more relevant, because the corona is a pure X-ray source and distinguishing between geometrically thin disc and corona would be much easier and unambiguous. This paper supports the corona interpretation at least in ZW229-15 because of positive detection or at least some presence indication of both optical break frequencies also in X-rays.

\subsubsection{Relativistic case}
\label{relativistic_case}

The model we have presented above postulates an origin for the short-term variability in the part of the accretion region far from the Innermost Stable Circular Orbit (ISCO), where relativistic effects are negligible. In the innermost regions of a magnetized accretion disc there could be different physical mechanisms which could induce instabilities. Here however we will consider whether the frequencies of the Rayleigh-Taylor (hereafter RT) modes, driven by the accretion of plasma inhomogeneities falling towards the accretion discs, could lie in the observed parameter range. Unfortunately, there exists very few fully relativistic treatments even of linearized instabilities relevant to our scenario. In the magnetically arrested disc model of accretion (\citealt{1976Ap&SS..42..401B}, \citealt{2011MNRAS.418L..79T}) plasma falling towards the accretion disc carries a magnetic field and accumulates at the interface between the corona and the disc before being accreted. This interface is thus potentially subject to RT instabilities. Recently, \citet{2016MNRAS.462..565C} have presented an analysis of linearized Rayleigh-Taylor instabilities in Kerr geometry which can be relevant to the case considered here. They suggest the following dispersion relation \citep[eq. 58]{2016MNRAS.462..565C}:

\begin{eqnarray}
\label{rt_disp}
\nu_{RT}^2 & = & -m^2\omega^2\nonumber\\
& + & m^2\left(\frac{M}{r^3}\right)\frac{r^6 \Delta}{A^2}\left[(1-c_s^4){\cal D}\left\{\rho\right\}
-(1+3c_s^2){\cal D}\left\{\frac{B^2}{8\pi}\right\}\right]\nonumber\\
& / & \left\{ \left[\frac{r-2M}{2\Delta A}[(6r^3-4Mr^2)a^2+(4r-9M)r^4]\right. \right. \nonumber\\
&& +\left. \frac{4M}{r}-\frac{5}{2}\right] {\cal D}\left\{\rho+p+\frac{B^2}{4\pi}\right\}\nonumber\\
&&\left. +\xi_{\rm K}\left(1-\frac{2M}{r}\right)\left(\rho+p+\frac{B^2}{4\pi}\right)\right\}
\end{eqnarray}

In the equation above $M, a$ are the BH's mass and spin, respectively, while $\rho, p, B$ are the (thin) accretion disc density, pressure and azimuthal magnetic field, respectively, and are assumed to depend only on the radial spatial coordinates $r$. This dispersion relation arises by assuming small perturbations of the form: $\delta(r)e^{\nu t + im\phi}$, and is restricted to the Kerr metric. All the remaining quantities ($\omega, \Delta, A, \xi_{K}$) are radial functions specific of the Kerr geometry, and they only depend on $M$ and $a$, while the symbol ${\cal D}$ denotes the jump of the quantity within the parentheses on the two sides of the interface: ${\cal D}\{ f \}= f_{+} - f_{-}$.

The relation above was obtained by \citet{2016MNRAS.462..565C} under a significant number of simplifying assumptions, for instance that the accretion disc's rotation is negligible. Despite that, it can be useful to estimate whether the characteristic timescales/frequencies of the unstable modes lie within the range permitted by the observations.

In the case of ZW 229-15 the BH mass has previously been established through reverberation mapping: $M = 1.00^{+0.19}_{-0.24} \times 10^{7}$ M$_{\sun}$ \citep{2011ApJ...732..121B}. For the accretion disc we assume a magnetized disc solution \citep{1986MNRAS.220..321K}, whose main parameters like the disc mass and central density are obtained by normalizing Kaburaki's profile to the luminosity of ZW 229-15.

\begin{figure}
\resizebox{\hsize}{!}{\includegraphics[angle=0]{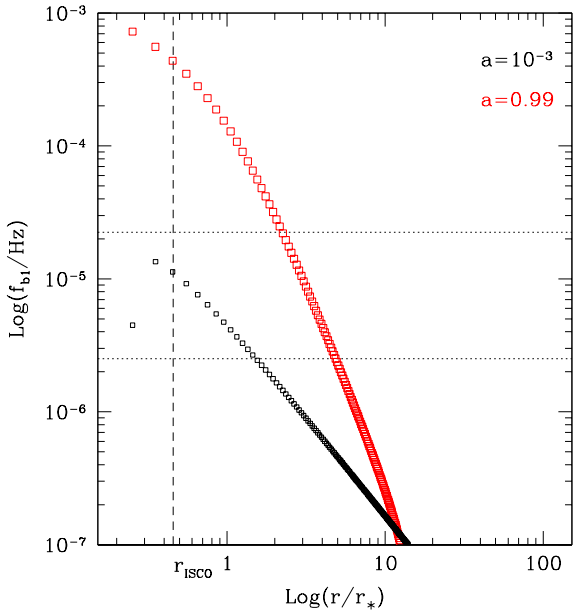}}
\caption{Frequencies of relativistic Rayleigh-Taylor instabilities. We plot the dispersion relation (eq. 58 from \citealt{2016MNRAS.462..565C}) for two extreme values of the ZW 229-15 spin $a$: $a = 10^{-3}$ (black squares) and $a = 0.99$ (red squares). Distances are in units of the Schwarzschild radius $r_{\ast}$. Note that the red squares trace \emph{imaginary} values of $\nu_{RT}$, corresponding to \emph{stable} modes. The horizontal band (from the lower limit ${\rm log}(f/{\rm Hz}) = -5.6$ of the x-axis in Fig.~\ref{pds_comp}, to the ${\rm log}(f_{\rm b2}/{\rm Hz}) = -4.65$ value) traces the allowed frequency region from the PDS analysis in Section~\ref{fits}; the vertical line marks the position of the Innermost Stable Circular Orbit, taken as the lower limit of the accretion disc.}
\label{fig_rt_disp}
\end{figure}

We plot in Figure~\ref{fig_rt_disp} the dispersion relation above for two extreme values of the spin. We see that for highly rotating BHs the only modes having frequencies comprised within the 5-days band are stable ($\nu_{RT}^{2} < 0$), while unstable solutions appear only for slowly rotating BH solutions. Solutions for intermediate values of $a$ can be either stable or unstable, depending on the distance, and lie between these two extreme cases. 

How can this dispersion relation be applied to explain the rapid variability observed in ZW 229-15 and other sources? We assume that unstable RT modes could give origin to the observed flickering: the question then is how these instabilities could arise. RT modes arise at the interfaces where density or other quantities (magnetic fields, pressure) are discontinuous. These interfaces can arise at the region separating the accretion disc and the corona when the accreting gas from the outside brings transient over-dense filamentary structures which can reach the accretion disc and locally increase the density at the interface \citep{1976ApJ...210..792A, 1976ApJ...207..914A, 1977ApJ...215..897E}. A detailed modeling would require numerical General Relativistic MHD simulations and spectral modeling, and is beyond the scope of the present paper. Yet, we note that the characteristic frequencies of these RT instabilities are consistent with those of the four AGNs considered in the previous paragraphs.

\subsection{\swift\ vs. \kepler\ data of ZW229-15}

The \swift\ data of ZW229-15 are too irregular to make any detailed cross correlation analysis with the \kepler\ data. However, the correlation is apparent in the upper right panel of Fig.~\ref{lc_swift}. The X-ray data show the same low flux plateau between day 735 and 750, the same declining trend below day 735 and rising trend above day 150 as the \kepler\ light curve. Moreover, the peaks around days 722, 727, 735, 757 and 765 are almost coincident. The main difference is the "continuum" behind/between consecutive peaks. While the \kepler\ flux smoothly declines between the mentioned peaks, the \swift\ flux experience a strong decline toward some "basic" level\footnote{Due to poor data coverage it is hard to say definitely whether this level is the same as the plateau level between days 735 and 750 or higher, but it seems to be slightly higher.} between consecutive peaks, i.e. the declining trend below day 735 seems to be composed by individual events with declining amplitude.

This suggests that only the peaks have origin in the both optical and X-rays, and the optical "continuum" is independent. The latter varies on a much longer time-scale than the former. A simple model where the longest variability is coming only from the outer accretion disc regions, and the faster is generated by the inner disc-corona duo can explain such a behaviour. The latter may be explained in two ways; 1) the mass accretion rate in the inner disc is feeding also the corona (by evaporation) and the fluctuations are present in both structures, which yields the same variability behaviour in both wave-bands, 2) the variability is generated by mass accretion fluctuations in the corona, and the optical signal is reprocessed by the underlying geometrically thin disc.

Moreover, the maximum flux of the X-ray peaks agrees with the optical trend/"continuum" on longer-time-scales. Note however that, based on the propagating fluctuation model (\citealt{lyubarskii1997}, \citealt{king2004}, \citealt{arevalo2006}) the mass accretion rate in the inner disc region is modulated by fluctuations generated further away in the outer disc regions. Therefore, the larger the mass in-flow from outer regions, the larger can be the fluctuations in the inner disc. This can generate correlated amplitudes of otherwise independent mass accretion fluctuations in the inner disc or evaporated corona.

\subsection{Rms-flux relation}

Generally speaking there is a relationship between the overall flux and the rms-flux data behaviour. All objects except KA1858+48 show some possible linear relations more evident at the lowest flux levels, with ZW229-15 being the most evident case. Probably the mean flux of the object plays a role as ZW229-15 and KA1858+48 have the largest and smallest fluxes, respectively.

Moreover, from Fig.~\ref{rms-flux_large} we see that in all the rms-flux relationships an initial linear relation is followed by a plateau or a slight depression. The same is also true for the \swift\ data and corresponding \kepler\ light curve interval of ZW229-15, showing possible correlation suggesting a common origin, and supporting the inner corona model. In \citet{dobrotka2016a} we showed that if a fast variability with linear rms-flux relation is superposed on an independent longer time-scale variability, the resulting rms-flux relation has very similar shape to the one presented in Figs.~\ref{rms-flux_large} and \ref{rms-flux_short}.

Therefore, we suppose that a variability with linear rms-flux relation is superposed on an independent signal with longer time-scale variability. This is valid for both the low (Fig.~\ref{rms-flux_large}) and high (Fig.~\ref{rms-flux_short}) resolution case of ZW229-15, yielding two linear rms-flux relations depending on how many light curve points are used for $\sigma_{\rm rms}$ and $\overline{\psi}$ calculation. In this picture an outer accretion disc generates propagating mass accretion fluctuations with long time-scales\footnote{Two estimates of the minimal limiting time-scale can be made. The value is either $\sim 4$ or $\sim 11.5$\,days. The former is based on 200 points with a cadence of 30\,min for $\sigma_{\rm rms}$ and $\overline{\psi}$ calculation in Fig.~\ref{rms-flux_large}, and the latter is based on a power rise for frequencies bellow ${\rm log}(f/{\rm Hz}) = -6$ in Figs.~\ref{pds_all} and \ref{pds_swift}.} behaving multiplicatively and showing the linear rms-flux relation for light curve with lower resolution in Fig.~\ref{rms-flux_large}. This linear rms-flux relation is deformed by underlying process generating even longer independent variability. In CVs this could for example arise from the mass feeding of the disc by the secondary star or superhump activity. In AGNs there are different physical processes which could result in rapid (from few hours to thousand of years) variations of the accretion disc's density. Recently \citet{cielo2017} have demonstrated that backflow of jet's gas originating from near the hotspot has a chaotic character, and can easily bring low angular momentum jet gas near the central accretion region. Although the simulations by \citet{cielo2017} do not have the spatial resolution to resolve the central accretion region and disc, they show that this backflow is a persistent feature during the first few million years since the launching of the jet. Another possible mechanism for large mass accretion rate variations on long time-scale is the radiation-pressure instability in the accretion disc (\citealt{wu2016}). This phenomenon generates some "heartbeat" oscillations or outbursts with a duration of two years in the case of a Seyfert galaxy IC 3599. The steep rise of the luminosity in the case of KA1925+50 or KA1904+37 could match this picture, but the variability/"flares" with time-scales of hundreds of days in the light curve of ZW229-15 are rather too short to be accounted for by this mechanism.

Further, the inner disc gas evaporates and forms the hot corona which follows every accretion rate fluctuation propagating inward from the outer disc regions, but the evaporated gas lives its own life as a turbulent region with its own characteristic time-scales of propagating mass accretion rate fluctuations. The former may be responsible for the X-ray amplitude correlation with the small peaks in the \kepler\ light curve, while the latter is responsible for the short-time-scale variability detected in optical and X-ray behaving also multiplicatively and manifesting as linear rms-flux trend in Fig.~\ref{rms-flux_short} being deformed by the larger variations.

This rms-flux shape explanation (if real) applies to the classical viscous scenario described in Section~\ref{classical_case}, but we do not know how it matches the relativistic scenario from Section~\ref{relativistic_case}. In the latter case, it is not sure whether the RT instability generates propagating mass accretion rate fluctuations behaving multiplicatively.

\subsection{Reality of the PDS features}

\kepler\ observatory is a very powerful tool for flickering study. However, there are some indications that the light curve analysis should be taken with caution. Some spurious excess power in the PDS can be present and the light curves should be reprocessed (e.g. \citealt{edelson2014}, \citealt{kasliwal2015a}, \citealt{kasliwal2015b}). Therefore, only a direct comparison of the \kepler\ PDSs with an independent source may yield satisfying conclusion. This has been done already by \citet{mushotzky2011} in the case of ZW229-15 on a long time-scale. Our comparison of optical and X-ray data offers an additional test. Besides the two frequency breaks in the PDS, it is also worth noting the rise of the PDS power towards lower frequencies from approximately ${\rm log}(f/{\rm Hz}) = -6$ observed in both optical and X-ray PDSs. Therefore, the credibility of the \kepler\ light curve is reinforced, at least in the case of ZW229-15.

\section{Summary and conclusions}

The results of this work can be summarized as follows:

(i) We discovered multiple component PDSs with two break frequencies of approximately ${\rm log}(f_{\rm b1}/{\rm Hz}) = -5.2$ and ${\rm log}(f_{\rm b2}/{\rm Hz}) = -4.7$ in \kepler\ data of four AGNs; ZW229-15, KA1925+50, KA1904+37 and KA1858+48.

(ii) For ZW229-15 $f_{\rm b1}$ is close to the 5\,day time-scale detected in the same data by \citet{edelson2014}, but probably it is a different PDS component.

(iii) We found indications of both break frequencies in X-ray \swift\ data of ZW229-15.

(iv) $f_{\rm b1}$ agrees with light-crossing or dynamical time-scale estimates of \citet{edelson2014}.

(v) $f_{\rm b2}$ agrees only with light-crossing scenario with time-scale estimated by \citet{edelson2014}, which requires an X-ray origin of the variability.

(vi) We found a viscous solution for $f_{\rm b1}$ and $f_{\rm b2}$, previously excluded by \citet{edelson2014}, if the disc is geometrically thick corona radiating in X-rays. The X-ray origin of fluctuations is supported by \swift\ data of ZW229-15. A further possibility is that relativistic unstable Rayleigh-Taylor modes could be operating in the innermost regions of the accretion disc.

(vii) All AGNs except KA1858+37 (perhaps because of low overall flux level) show rising rms-flux relation with possible linear shape. This relation is not valid for whole flux interval. After an initial (linear) rise a plateau, slight depression and/or subsequent rise follows. Based on previous simulations by \citet{dobrotka2016a} we suggest that such rms-flux deformed shape (initially linear) is generated by an independent underlying modulation with longer/longest time-scale. Their physical origin is not clear, but various plausible scenarios like backflow of jet's gas or radiation-pressure instability are possible.

(viii) The same deformed rms-flux shape with initial potentially linear rising trend shows also the X-ray \swift\ data of ZW229-15 on much shorter time-scale. Optical \kepler\ data do show similar potentially correlated trend on similar short-term time-scale suggesting common origin, i.e. the central geometrically thick hot X-ray corona. The deformed shape agrees with previous simulations suggesting that the short-term variability is superposed on longer independent variability (see previous point).

(ix) The linearity of the rms-flux data, or the interval where it has the linear shape, depends on the time-scale (number of light curve points) for rms-flux data calculation. The existence of a short-term and long-term time-scale solutions suggests the existence of two superposed independent variabilities, both having the multiplicative character typical for the accretion rate fluctuation propagation model. The longer time-scale variability can be generated by the whole accretion disc, while the shorter can be generated by the evaporated gas at the inner disc forming the inner hot corona. 

\section*{Acknowledgment}

AD was supported by the Slovak grant VEGA 1/0335/16 and by the ERDF - Research and Development Operational Programme under the project "University Scientific Park Campus MTF STU - CAMBO" ITMS: 26220220179.

\bibliographystyle{mn2e}
\bibliography{mybib}

\label{lastpage}

\end{document}